\documentclass[aps,prl,reprint,superscriptaddress]{revtex4-1}

\usepackage{graphicx}
\usepackage{amsmath}
\usepackage[separate-uncertainty]{siunitx}
\usepackage{xcolor}

\graphicspath{{figures/}}

\newcommand*{\ngml}{\nano\gram\per\milli\liter}
\newcommand*{\ugml}{\micro\gram\per\milli\liter}
\newcommand*{\mgml}{\milli\gram\per\milli\liter}

\begin{document}

\author{Kaitlynn M. Snyder}
\altaffiliation{These authors contributed equally.}
\affiliation{Department of Physics and Center for Soft Matter Research,
  New York University, New York, NY 10003, USA}

\author{Rushna Quddus}
\altaffiliation{These authors contributed equally.}
\affiliation{Department of Chemistry, New York University, New York, 
  NY 10003, USA}

\author{Andrew D.\ Hollingsworth}
\affiliation{Department of Physics and Center for Soft Matter Research, 
  New York University, New York, NY 10003, USA}

\author{Kent Kirshenbaum}
\affiliation{Department of Chemistry, New York University, New York, 
  NY 10003, USA}

\author{David G. Grier}
\email{david.grier@nyu.edu}
\affiliation{Department of Physics and Center for Soft Matter Research, 
  New York University, New York, NY 10003, USA}

\title{Holographic Immunoassays}

\begin{abstract}
 The size of a probe bead reported by holographic particle
  characterization depends on the proportion of the surface
  area covered by bound target molecules and so can
  be used as an assay for molecular binding.
  We validate this technique by measuring
  the kinetics of irreversible binding for the antibodies
  immunoglobulin G (IgG) and immunoglobulin M (IgM) 
  as they attach
  to micrometer-diameter colloidal
  beads coated with protein A.
  These measurements yield the antibodies'
  binding rates and can be inverted
  to obtain the concentration of antibodies in solution.
  Holographic molecular binding assays therefore
  can be used to perform fast quantitative immunoassays
  that are complementary to conventional
  serological tests.
\end{abstract}

\maketitle

\section{Introduction: Holographic molecular binding assays}
\label{sec:HPC}

\begin{figure*}
    \centering
    \includegraphics[width=0.75\textwidth]{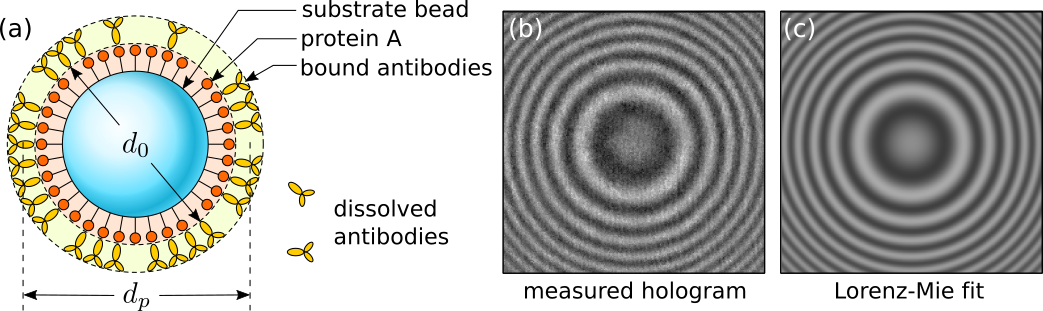}
    \caption{(a) 
    Schematic representation of a molecular binding assay based on holographic particle characterization. Probe beads consist of
    spherical polystyrene substrates coated with functional groups (protein A)
    that can bind target antibodies
    from solution.
    The probe beads have an effective diameter
    that increases from $d_0$ to $d_p$ when
    antibodies bind.
    (b) A molecular-scale coating of antibodies
    influences the recorded hologram of a bead.
    (c) This change can be quantified by fitting
    to predictions of the Lorenz-Mie theory of
    light scattering, yielding an estimate for
    the fractional surface coverage and from this the
    concentration of antibodies.}
    \label{fig:schematic}
\end{figure*}

Holographic molecular binding assays use
holographic particle characterization
\cite{lee2007characterizing}
to directly measure changes in the diameters of
micrometer-scale colloidal spheres
caused by molecules binding to their surfaces
\cite{cheong2009flow,zagzag2020holographic}.
This rapid measurement technique
eliminates the need for
fluorescent labeling to detect binding and thus
reduces the cost, complexity, time to completion
and expertise required for standard binding
assays such as ELISA. 
Being based on batch-synthesized beads,
holographic molecular binding assays
do not require microfabricated sensors and can
be performed with comparatively little sample
preparation.
Holographic molecular binding assays therefore
have great promise as
medical diagnostic tests, particularly the
serological tests required to assess patients'
immune responses to pathogens such as
SARS-CoV-2, the coronavirus responsible for
COVID-19.

The ability to measure nanometer-scale changes in the
diameters of micrometer-scale spheres is provided 
by quantitative analysis of single-particle holograms
obtained with in-line holographic video microscopy
\cite{sheng2006digital,lee2007characterizing}.
The hologram of an individual colloidal sphere is
fit to a generative model based on the Lorenz-Mie theory of light 
scattering 
\cite{mishchenko2002scattering,bohren2008absorption,gouesbet2011generalized}
to extract the particle's
diameter, $d_p$, refractive index, $n_p$
and three-dimensional position, $\vec{r}_p$ \cite{lee2007characterizing}.
This measurement scheme is depicted schematically in Fig.~\ref{fig:schematic}.
One such measurement can be completed in a few
milliseconds and yields a bead's diameter
with a precision of \SI{5}{\nm} and its
refractive index to within \SI{1}{ppt}.
A set of such measurements can be used to measure
the mean diameter of a population of particles
to within a fraction of a nanometer
\cite{zagzag2020holographic}, which is
sufficient to detect the growth of
molecular-scale coatings.

Previous demonstrations of holographic molecular binding assays 
\cite{cheong2009flow,zagzag2020holographic} have
reported changes in probe beads' properties when
the concentration of target molecules is large enough
to saturate the beads' binding sites.
Here, we report concentration-dependent
trends that cover
the range from zero analyte to binding-site 
saturation.
Interpreting these results through the statistical mechanics
of molecular binding then achieves three goals:
(1) to use holographic binding assays to probe the
kinetics of molecular binding;
(2) to validate the effective-sphere model used to
interpret holographic particle characterization measurements
on coated spheres; and
(3) to establish the effective range of
analyte concentrations over which holographic binding
assays can quantitate target molecules in
solution, a key capability for clinical testing.

\section{Experimental}

We demonstrate quantitative
holographic binding assays through measurements on antibodies
binding to beads coated with protein A,
specifically
immunoglobulin G (IgG) and immunoglobulin M (IgM).
These are
well-studied model systems \cite{lund2011exploring}
with which to validate holographic binding assays
and to establish their detection limits.
Given the central role of IgG and IgM in the immune 
response to viral pathogens, these 
experimental demonstrations 
furthermore serve as models for fast,
inexpensive and quantitative serological tests.

\subsection{Probe beads and buffer solution}

The probe beads used for this study
(Bangs Laboratories, catalog no.~CP02000,
lot no.~14540)
have a polystyrene core with a nominal diameter of
$d_0 = \SI{1}{\um}$
and  a surface layer of immobilized protein A molecules, each of which
has five
binding sites for the Fc region of immunoglobulins
\cite{deisenhofer1981crystallographic,moks1986staphylococcal}.
These functionalized beads are dispersed at a concentration of
\SI{2e6}{particles\per\milli\liter} in
an antibody binding buffer.
The same buffer is used to dissolve antibodies
for testing.
Equal volumes of the probe-bead dispersion and the
antibody solution are mixed to initiate
incubation.

The antibody binding buffer consists of
\SI{50}{\milli M} sodium borate buffer prepared
with boric acid 
(\SI{99.5}{\percent}, Sigma-Aldrich, 
catalog no.~B0394, lot no.~SLBM4465V) and NaOH (\SI{98}{\percent}, Sigma-Aldrich, catalog no.~S8045, lot no.~091M01421V) in deionized water
(\SI{18.2}{\mega\ohm\centi\meter}, Barnstead
Millipure).
The pH of the buffer is adjusted to \num{8.2}
with the addition of dilute HCl
(\SI{38}{\percent}, Sigma-Aldrich, 
catalog no.~H1758) to optimize the binding
of antibodies to protein A
\cite{fishman2019protein}.

The dispersion of functionalized colloidal
spheres constitutes a bead-based assay kit
for immunoglobulins that bind to protein A.
The same approach can be used to create
specific immunoassays for particular antibodies
by functionalizing the beads' surfaces 
with suitable antigens instead of protein A.
Multiplexed assays can be produced by
separately functionalizing substrate beads
that can be distinguished holographically
by size or by refractive index and then mixing
their dispersions to make a test kit.

\subsection{Assay protocol}

An assay is performed by dissolving 
target antibodies
in the buffer at concentrations 
from \SI{200}{\ngml} up to
\SI{200}{\ugml}.
Antibody solution is then mixed with
an equal volume of 
the stock dispersion of probe beads
to obtain a bead concentration
of \SI{e6}{particles\per\milli\liter}
and antibody concentrations in the
range from \SI{100}{\ngml} to \SI{100}{\ugml}.
This easily allows for detection in a physiologically relevant range following suitable dilution, as
the typical concentration of immunoglobulins
in human serum is \SI{10}{\mgml}
\cite{cassidy1975human}.
The sample is allowed to equilibrate for
$\tau = \SI{45}{\minute}$ at room temperature before
being analyzed.

To model immunoassays that would be relevant
for serological testing,
we performed assays on 
rabbit IgG 
(EMD Millipore; catalog no.~PP64, lot no.~3053798)
and
human IgM
(Sigma-Aldrich; catalog no.~I8260, lot no.~069M4838V).
Aggregation of IgM is suppressed by increasing
the ionic strength of the buffer 
through the addition of 
\SI{150}{\milli M} of NaCl
(\SI{99.5}{\percent}, Sigma-Aldrich,
catalog no.~S7653) \cite{fishman2019antibody}.

Control measurements are performed by replacing
the antibodies with alcohol dehydrogenase
(ADH, Sigma-Aldrich; catalog no.~A3263-7.5KU, lot no.~SLBW31382).
Non-specific binding due to incomplete
coverage of the bead surfaces by
protein A is blocked for these experiments by incubating
the probe beads with bovine serum albumin
(BSA, Sigma-Aldrich, catalog no.~A2153).
BSA adsorbs non-specifically to exposed 
polystyrene
and does not interfere with antibody 
binding to protein A.
ADH does not bind to either protein A or BSA
and thus should not attach to the probe beads.
With a molecular weight greater than
\SI{140}{\kilo\dalton}, ADH is comparable
in size to IgG and thus should have
a similar holographic signature, were it to
bind.

\subsection{Holographic particle characterization}

Holographic particle characterization measurements
are performed with a commercial 
holographic particle characterization
instrument
(Spheryx xSight) 
set to record holograms at a wavelength of
\SI{447}{\nm}.
Each measurement involves pipetting a
\SI{30}{\micro\liter} aliquot of the dispersion
into the sample reservoir of one channel 
in an eight-channel microfluidic chip (Spheryx xCell).
The sample chip is then loaded into xSight,
which is set to draw \SI{1}{\micro\liter}
of the sample through the observation volume
in a pressure-driven flow with a peak speed
around \SI{3}{\mm\per\second}.
Data for a thousand beads is collected
in measurement time
$\Delta\tau = \SI{2}{\minute}$ and is fully
analyzed in about \SI{15}{\minute}.

The Lorenz-Mie theory used to analyze holograms
treats each particle as a homogeneous sphere.
When applied to inhomogeneous particles, such as
the coated spheres in the present study, the
extracted parameters must be interpreted as
representing the properties of an effective sphere
\cite{cheong2011holographic,odete2020role,altman2020interpreting}.
These effective-sphere properties will differ
from the physical properties of the coated sphere
unless the coating has the same refractive index
as the substrate bead.
The refractive index of the coating, moreover,
depends on the fraction, $f$, of binding sites
occupied by molecules, which means that the effective
diameter of the coated sphere also depends on $f$.
Numerical studies show that the holographically
measured diameter increases linearly with
surface coverage \cite{altman2020interpreting},
\begin{equation}
  \label{eq:diameter}
  d_p = d_0 + 2 \delta \, f,
\end{equation}
where $d_0$ is the bare sphere's diameter and
$\delta$ is the effective optical thickness
of a complete layer of bound molecules.
The value of $\delta$ depends on
the size of the target molecule,
the density of binding sites,
and the refractive index of the target molecule 
relative to those of the medium and the substrate bead 
\cite{altman2020interpreting}.

\begin{figure*}
    \centering
    \includegraphics[width=0.9\textwidth]{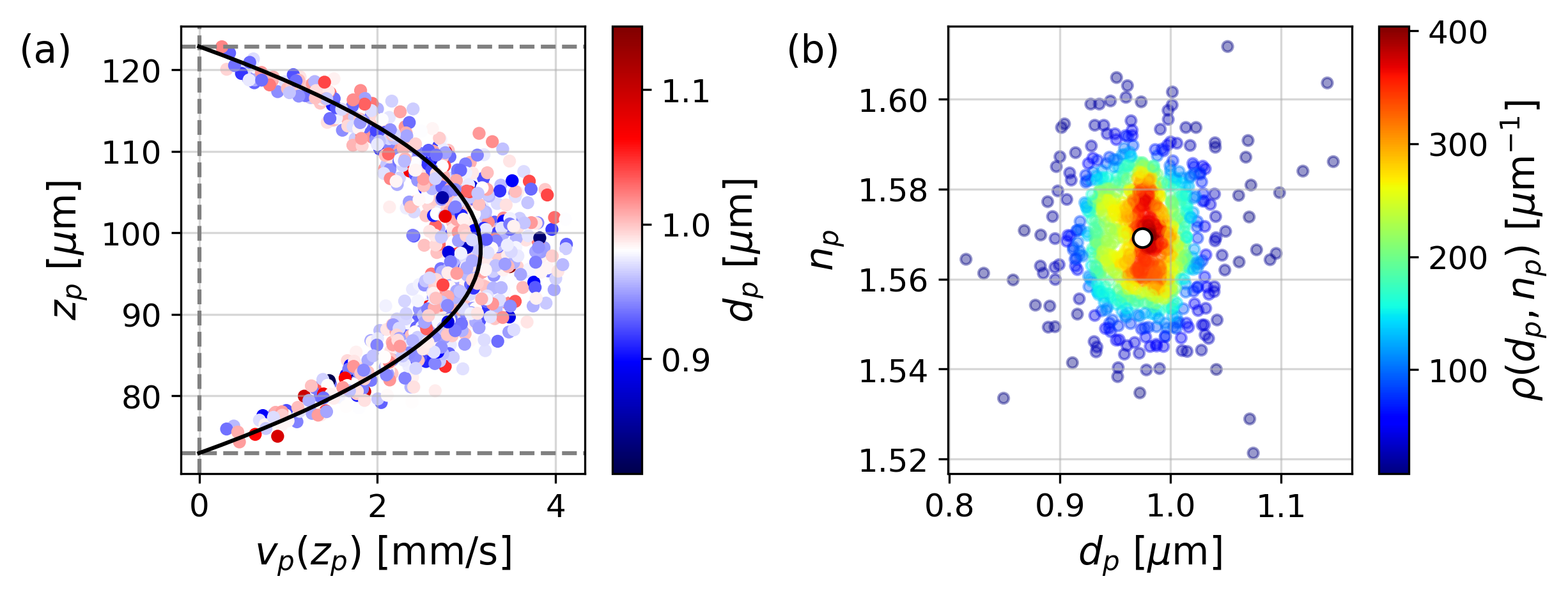}
    \caption{Typical holographic molecular binding assay
    for a sample of probe beads incubated with
    \SI{10}{\micro\gram\per\milli\liter} IgG.
    (a) Holographically measured velocity profile. Each point represents the speed, $v_p$, of a single bead
    as a function of its axial position, $z_p$,
    relative to the instrument's focal plane.
    The solid curve is a fit to the parabolic 
    Poiseuille flow profile.
    Horizontal dashed lines indicate the axial positions
    of the channel's walls inferred from this fit.
    Points are colored by each particle's measured 
    diameter, $d_p$. Evenly mixed colors
    demonstrate that the results are not biased by the
    particles' positions in the channel.
    (b) Holographic characterization data for the same
    sample of beads showing the distribution of 
    single-particle diameter, $d_p$, and refractive
    index, $n_p$. Points are colored by the density
    of measurements, $\rho(d_p, n_p)$. The central
    point shows the population mean for this sample
    and is sized to represent the uncertainty in the mean.
    }
    \label{fig:example}
\end{figure*}

Each dispersed particle is recorded and
analyzed up to \num{10} times as it traverses
the observation volume and the resulting
three-dimensional position measurements
are linked into a trajectory 
\cite{crocker1996methods}.
The linked measurements are
combined to improve
the precision of the estimated values for
the particle's diameter and refractive index
\cite{cheong2009flow}.
Typical results for a sample of beads
incubated with \SI{10}{\ugml} of IgG
are presented in Fig.~\ref{fig:example}.
Each point in these scatter plots represents
the holographically measured
trajectory, Fig.~\ref{fig:example}(a),
and properties, 
Fig.~\ref{fig:example}(b),
of a single particle.
The size of the dots is comparable to the
estimated measurement precision.

Single particle trajectories are useful for
mapping the fluid flow
in the microfluidic channel. \cite{cheong2009flow}
Figure~\ref{fig:example}(a) shows the
beads' speed, $v_p(z_p)$, 
as a function of axial position, $z_p$,
relative to the instrument's focal plane.
Data points in Fig.~\ref{fig:example}(a) are
colored by the spheres' measured diameters
and show that particles are distributed uniformly
throughout the channel and that particle size
is not correlated with height in the channel.
Fitting these data to the anticipated parabolic Poiseuille
flow profile yields
estimates for the positions of the upper and
lower walls of the channel, which are
indicated by the horizontal dashed lines in Fig.~\ref{fig:example}(a).
Mapping the flow profile provides an important
internal reliability check, ensuring that the 
sample has flowed smoothly through the channel,
that the microfluidic channel is properly seated in the
instrument, and that trajectory linking has
proceeded correctly.

Figure~\ref{fig:example}(b) show the single-particle
characterization data obtained from these trajectories,
with each point representing the effective diameter, $d_p$,
and refractive index, $n_p$, of a single bead.
Plot symbols are colored by the
density of observations, $\rho(d_p, n_p)$.

The \num{890} particles in this
data set enable us to compute the 
population-average diameter,
$d_p = \SI{0.974(2)}{\um}$
and the mean refractive index,
$n_p = \num{1.5692(8)}$.
The value for the refractive index
is significantly smaller than
the value of \num{1.60} expected for polystyrene
at the imaging wavelength, and is
consistent with expectations
for a coated sphere in the
effective-sphere interpretation
\cite{altman2020interpreting}.
The mean diameter is significantly
larger than the baseline value of
$d_0 = \SI{0.964(2)}{\um}$
obtained for the probe beads alone.
The difference, 
$\Delta_p = d_p - d_0 = \SI{10(3)}{\nm}$
is consistent with
a statistically significant
detection of antibody binding
\cite{zagzag2020holographic}
at concentrations orders of magnitude
lower than physiological levels
\cite{cassidy1975human,goldstein2006selective,long2020antibody}.

A principal aim of the present study is to
combine the effective-sphere analysis
of probe beads' holograms \cite{cheong2009flow,zagzag2020holographic,altman2020interpreting}
with the statistical physics of molecular binding
to obtain quantitative information
on the kinetics of antibody binding
from measurements of $d_p(c,t)$.
Conversely, this analysis
establishes that a holographically observed
shift in bead diameter can be used
to measure the
concentration of antibodies in solution
and furthermore 
establishes the trade-off between
concentration sensitivity and
measurement time for such
holographic immunoassays.

\subsection{Kinetics of molecular binding}

Antibodies bind
rapidly to protein A in the antibody binding buffer
and the rate of dissociation is small enough
for the process to be considered irreversible
\cite{norde1992energy}.
Antibodies therefore continue to 
bind to the probe beads until 
all of the
surface sites are saturated
or the solution is depleted.
Assuming that depletion may be ignored
and the solution remains well mixed, the fraction of occupied
sites, $f(c, t)$, increases at a rate that
depends on the concentration of antibodies, $c$, 
and the availability of unoccupied sites
\cite{privman1991continuum,buijs1996adsorption,adamczyk2000kinetics}
\begin{equation}
    \label{eq:ritsi}
    \frac{df}{dt} = \gamma(c) [1 - f(c, t)].
\end{equation}
This model differs from those in previous studies
\cite{dancil1999porous,ogi2007concentration,nelson2015mechanism}
by not having to account for detachment of antibodies 
from binding sites.
Minimizing unbinding optimizes the sensitivity of
the assay to small concentrations of analyte 
and reduces the time required to perform
measurements.

The rate constant, $\gamma(c)$, accounts
for the microscopic kinetics of molecular
binding.
Further assuming that the concentration of antibodies
is low enough that binding events are independent,
we model $\gamma(c) = k c$, where $k$ is the binding rate for the antibodies in the antibody binding buffer.
The solution to Eq.~\eqref{eq:ritsi},
\begin{equation}
\label{eq:f}
    f(c, t) = 1 - e^{- k c t},
\end{equation}
satisfies the initial condition $f(c, 0) = 0$ and
shows that binding assays can be performed
either as a function of time for fixed antibody 
concentration, $c$, or as a function of concentration
at fixed incubation time, $t$.
If, furthermore, the measurement is performed
over a time interval, $\Delta\tau$, starting
after incubation time $\tau$, the average
coverage is
\begin{subequations}
\label{eq:fbar}
\begin{align}
    \bar{f}(c, \tau)
    & = 
    \frac{1}{\Delta\tau}
    \int_\tau^{\tau+\Delta\tau} f(c, t) \, dt \\
    & = 
    1 - 
    \frac{1 - e^{-k c \Delta \tau}}{k c \Delta\tau} \,
    e^{- k c \tau}.
\end{align}
\end{subequations}

\subsection{Monitoring binding holographically}
\label{sec:diameterincrease}

Combining Eq.~\eqref{eq:diameter} with
Eq.~\eqref{eq:fbar}
yields an expression for the dependence of
the measured bead diameter on the target molecules'
concentration in solution:
\begin{equation}
  \label{eq:diametershift}
  \Delta_d(c, \tau) \equiv d_p - d_0
  =
  2 \delta \,
  \left(
    1 - 
    \frac{1 - e^{-k c \Delta \tau}}{k c \Delta\tau} \,
    e^{- k c \tau}
  \right).
\end{equation}
Holographic measurements of $\Delta_d(c,\tau)$ 
at fixed incubation time $\tau$ can be
interpreted with Eq.~\eqref{eq:diametershift}
to estimate the effective layer thickness, $\delta$,
and the rate constant, $k$.
These values, in turn, can be used to anticipate
how the sensitivity of the assay for
antibody concentration depends on 
incubation time, $\tau$.
This sensitivity can be further improved by reducing
uncertainties in $\Delta_d(c, \tau)$, either
by extending the measurement time to analyze more beads
or by optimizing the optical properties of the beads
to increase $\delta$
\cite{altman2020interpreting}.

\begin{figure*}
  \centering
  \includegraphics[width=0.9\textwidth]{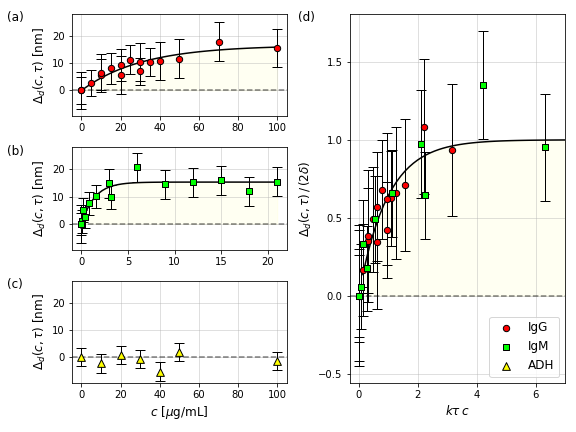}
  \caption{Holographic molecular binding assays
  for (a) IgG (red circles) and (b) IgM (green squares)
    to colloidal beads coated with protein A
    dispersed in antibody binding buffer.
    IgM assay is performed with \SI{150}{\milli M}
    added NaCl to suppress aggregation.
    Discrete points show the increase,
    $\Delta_d(c, \tau) = d_p(c, \tau) - d_0$, 
    of the population-average
    effective-sphere diameter, $d_p(c, \tau)$, 
    relative to the probe beads'
    reference diameter, $d_0$, as a function
    of antibody concentration, $c$
    after fixed incubation time 
    $\tau = \SI{45}{\minute}$.
    Solid curves are best-fits to
    Eq.~\eqref{eq:diametershift}
    for measurement time $\Delta\tau = \SI{2}{\minute}$.
    (c) 45~minute incubation with alcohol
    dehydrogenase (ADH) has no measurable
    affect on probe bead diameters.
    (d) Binding data collapsed according to Eq.~\eqref{eq:diametershift}.
    Concentrations are scaled by $k \tau$ and 
    diameter shifts are scaled
    by the layer thickness, $\delta$.}
  \label{fig:experiment}
\end{figure*}

\section{Results}

The discrete points in Fig.~\ref{fig:experiment}
show measured shifts, $\Delta_d(c, \tau)$, in
the population-average bead diameter after
$\tau = \SI{45}{\minute}$ incubation with
(a) IgG, (b) IgM and (c) ADH.
These shifts are measured in nanometers and
illustrate the precision with which
holographic particle characterization can
resolve the diameters of probe beads.
Error bars indicate uncertainties in the
mean diameter given the particle-counting
statistics for each measurement. A single
point represents results from roughly
\num{1000} beads observed
in \SI{1}{\micro\liter} of the sample
over $\Delta \tau = \SI{2}{\minute}$.

As anticipated, bead diameters increase
upon incubation with antibodies
by an amount that depends on
antibody concentration.
Incubation with ADH has no such effect
presumably because ADH does not bind
to protein A.
Results for IgG and ADH are presented
for concentrations up to \SI{100}{\ugml}.
IgM is plotted only up to \SI{20}{\ugml}
because $\Delta_d(c,t)$ reaches 
a plateau beyond $c = \SI{5}{\ugml}$,
which we interpret to represent saturation
of the available surface sites by IgM.

The solid curves in
Fig.~\ref{fig:experiment}(a) and 
Fig.~\ref{fig:experiment}(b)
are fits of the measured bead diameters
to Eq.~\eqref{eq:diametershift} for
the apparent layer thickness, $\delta$,
and the rate constant, $k$.
Interestingly, fits to the data for
both IgG and IgM
are consistent with an effective layer thickness of
$\delta = \SI{8.0(5)}{\nm}$ even though
IgM has five times the molecular weight of IgG.
This agreement could be a coincidence
arising from the effective-sphere interpretation
of holographic imaging data \cite{altman2020interpreting}.
It also is consistent with a model
in which multi-site binding of the predominantly pentameric IgM assembly results in a flattened orientation of the IgM on the probe beads'
surfaces, thus contributing no more
to $\delta$ than the single domain of IgG.


The fit value for the rate constant of
IgG is 
$k_\text{G} = 
\SI{1.2(3)e-5}{\milli\liter\per\micro\gram\per\second}$,
which corresponds to a rate per binding site of
$m_\text{G} k_\text{G}
= 
\SI{3.0(8)e-18}{\milli\liter\per\second}$,
given the $m_\text{G} = \SI{150}{\kilo\dalton}$
molecular weight of IgG.
We express this figure as a rate of binding
events per surface site
rather than as a rate per molecule to emphasize that
the molecules are filling available binding sites
on the probe beads.


The corresponding rate constant for IgM,
$k_\text{M} = 
\SI{2.5(8)e-4}{\milli\liter\per\micro\gram\per\second}$,
is an order of magnitude greater than $k_\text{G}$.
The difference becomes even greater
when
account is taken of the
$m_\text{M} = \SI{970}{\kilo\dalton}$
molecular mass of pentameric IgM:
$m_\text{M} k_{M}
= 
\SI{4.1(12)e-16}{\milli\liter\per\second}$.
Naively assuming that each 
IgG molecules occupies $\nu_\text{G} = 1$
binding site and each IgM occupies
$\nu_\text{M} = \num{5}$ 
reduces the difference proportionately,
\begin{equation}
    \frac{m_\text{M} k_\text{M}}{\nu_\text{M}}
    \frac{\nu_\text{G}}{m_\text{G} k_\text{G}}
    =
    \num{27(1)}.
\end{equation}
The remaining large difference in binding rates
cannot be ascribed to differences in
bulk transport properties because the
molecules' diffusion constants are proportional
to their sizes, which suggests that IgG
should attach more rapidly, being smaller.
It may instead reflect differences 
in the two antibodies' microscopic
binding mechanisms \cite{law2019igm}.
Possible explanations include differences
in binding probabilities as molecules approach the surface due to the multivalent presentation of binding sites for the pentameric IgM. In addition, different barriers to attachment may arise due to variations in the nature of electrostatic interactions for immunoglobulins. A more thorough evaluation of the influence of multivalency on attachment kinetics for IgGs, IgMs and other biomacromolecules will provide an intriguing challenge for our future studies. Nevertheless, even a simplified model such as a one-to-one binding mode between Protein A and IgG, for example, provides the capability to conduct chemical analysis of immunoglobulin concentration in solution.

Given our primary goal of developing
immunoassays for serological testing,
the experimental results in 
Fig.~\ref{fig:experiment}
confirm that holographic particle characterization 
provides a basis for quantitative measurements
of antibody concentrations under physiological
conditions.
The success of these fits to a kinetic
model for attachment is demonstrated
by the data collapse in Fig.~\ref{fig:experiment}(d),
with results from IgG and IgM both falling
on the same master curve despite the
substantial difference in the two antibodies'
rate constants.

\section{Conclusion}

This study has demonstrated that holographic
particle characterization can perform
quantitative molecular binding assays,
including measuring the rate constants that
characterize molecular binding.
Our results demonstrate that a single
\SI{15}{\minute} measurement can quantify
the concentration of IgG in solution down to
concentrations as low as 
\SI{10}{\micro\gram\per\milli\liter}
and concentrations of IgM as low as
\SI{1}{\micro\gram\per\milli\liter}.
Longer measurements and 
larger statistical samples can
improve this sensitivity, both by
increasing occupancy of binding sites
and also by reducing uncertainty in
the diameter shift.

Whereas the IgG-protein A system has been
studied extensively, less has been written
about binding of IgM to substrates coated
with protein A.
The holographic assays reported here 
provide insights into the binding mechanism
that may inform future studies.
We find, for example, that IgM tends to bind significantly more 
rapidly to protein A than IgG. 
Our observations also suggest
that IgM may tend to bind flat to the surface of
a functionalized bead.
How these trends depend on such factors as
electrolyte composition and concentration
fall outside the intended scope of the present
study and will be addressed elsewhere.

Using protein A to provide binding
functionalization yields a general-purpose
assay for antibody concentration, rather than an
immunoassay for specific antibodies.
This general-purpose assay already should
be useful as a rapid screening test for
Antibody Deficiency Disorders
\cite{ballow2002primary,patel2019expanding}.

Holographic immunassays can be targeted
for specific diseases
by replacing protein A
as a surface binding group with
appropriate specific antigens, including peptides, proteins, or other biomolecules.
Such functionalized colloidal spheres are standard
components of conventional bead-based
assays, which typically rely on fluorescent
labels for readout.
Holographic analysis
yields results faster and at lower cost
by eliminating reagents, processing
steps and expertise 
needed to apply fluorescent labels.
Holographic analysis furthermore
yields quantitative results for antibody
concentration without requiring 
extensive calibration.
The speed and sensitivity of holographic immunoassays
can be improved further by optimizing the sizes
and optical properties of the substrate beads.
Such efforts currently are under way.

\section*{Acknowledgments}

This work was supported by the RAPID program of the
National Science Foundation under Award No.\ DMR-2027013.
Partial support was provided by NASA through Grant
No.\ NNX13AR67G.
The Spheryx xSight holographic characterization instrument used in this study was acquired by New York University's 
Materials Research Science and Engineering Center as shared instrumentation
with support from the NSF under 
Award No.\ DMR-1420073.

%

\end{document}